# Time-resolved investigations of a glow mode impulse dielectric barrier discharge in pure ammonia gas by means of E-FISH diagnostic

R. JEAN-MARIE-DESIREE[1], A. NAJAH[1], C. NOËL[1], L. De POUCQUES[1], S. CUYNET[1]

[1] Université de Lorraine, CNRS, IJL, F-54000 Nancy, France

**Abstract:** Time-resolved electric field strength measurements have been performed, using an electric-field induced second harmonic (E-FISH) diagnostic, in a nanosecond glow discharge of an impulse dielectric barrier discharge (iDBD), in an ammonia gas environment. A temporal resolution of 2 ns and a spatial resolution estimated at 70 µm (given by laser waist) have been achieved. The comparative study of E-FISH measurements with and without a plasma discharge, operated at 4 kHz, reveal the presence of a persistent counter electric field, which is assumed to be caused by charge accumulation in between the AlN dielectrics used. Furthermore, by studying the influence of the applied voltage, the pressure, and the inter-dielectric distance, measurements seem to indicate the presence of charges remaining also in the post-discharge volume from the previous discharge to the next one.

## 1. Introduction

Over the last decades, non-thermal plasma developed with dielectric barrier discharge (DBD) cells has been extensively studied, for various fields such as biomedical applications [1–3], chemical fuel reforming processes [4,5], plasma actuators for high-speed flow control [6] or icing mitigation [7] and surface modifications [8–11]. These discharges expose three different operating modes [8,12–15]: filamentary, diffuse (or also qualified as homogeneous) and patterned (or also named as self-organized). The filamentary discharge mode is characterized by a multitude of streamers, randomly distributed in time and space over the dielectric surface. On the other hand, the diffuse discharge mode is initiated by a Townsend breakdown mechanism that extends over the entire electrode surface with a low ionization degree compared to the filamentary mode. Additionally, Massines and coworkers [16,17] reported different mechanisms responsible for sustaining such a diffuse discharge in different gas mixtures. Therefore, they proposed a classification of atmospheric diffuse dielectric barrier discharges into two main categories. One, named atmospheric pressure Townsend discharge (APTD), can be characterized by a plasma emission localized close to the anode and quasi-constant value of the electric field across the discharge gap. The second one is called atmospheric pressure glow discharge (APGD), derived from low-pressure glow discharge, as its behavior exhibits similarities with the latter (*e.g.* cathode fall, negative glow, Faraday dark space, *etc.*). Henceforth, the electric field across the discharge gap in an APGD is not constant. Finally, the patterned mode results in localized and reproducible microdischarges, also initiated by Townsend mechanism.

Indeed, the need of spatial and temporal accurate diagnostics for the electron density and temperature, species energy and density are fundamental for a better understanding of the chemical and physical processes occurring into theses plasmas. As the charged particles are driven by the electric field, considerable efforts have been made in order to assess its space and time distributions using intrusive method as electrostatic fluxmeters [18], calibrated capacitive probe [19] or Pockels electro-optical crystals [20,21]. However, the introduction of such devices disturbs the discharge, due to the alteration of the power circuit impedance, or surface charge accumulation, and modification, which may occur. Hence, the introduction of non-intrusive methods based on optical measurements [22], including optical emission spectroscopy [23,24], coherent anti-Stokes Raman spectroscopy [25–27] have been recently developed. Nonetheless, since both of these methods rely on resonant mechanisms, they are species-dependent and cannot be used for a broad range of gases.

In the recent years, a brand-new non-intrusive optical method has been under development, working with the second harmonic generation from a laser-excited medium. This optical non-linear effect was first experimentally demonstrated by Franken *et al.* [28] (1961) and used to measure arbitrarily applied electric fields by Dogariu *et al.* [29] for the first time. Indeed, the intensity of the second harmonic radiation is found to be proportional to the square of an external electric field, as it affects the polarizability of the medium. Since its inception, this electric-field induced second harmonic (E-FISH) technique has been implemented to perform time-resolved electric-field measurements in surface DBD [30,31], in nanosecond pulsed DBD [23,32–34] and for understanding fast ionization wave propagation. Since this diagnostic is based on radiative phenomena, the decay time mentioned to be about several hundred of picosecond enables such rapid kinetics to be characterized. Moreover, the E-FISH diagnostic can be performed on any species using the same fundamental laser beam. This technique is sensitive to the electric field direction [33] with a spatial and temporal resolutions conditioned by the laser probe.

Finally, the E-FISH diagnostic allows a better understanding of rapid charge transport and ionization wave, before, during and after the discharge. The aim of the present work is then to measure the axial electric field strength in a glow nanosecond discharge, from an impulse dielectric barrier discharge (iDBD), in an ammonia gas environment, by employing the E-FISH diagnostic. The iDBD cell, in a $1.0 \times 10^4$ Pa ammonia gas, has already been used in order to functionalize an organic ligand involved in a metal-organic framework [9]. To gain insight into this glow iDBD [17] in pure ammonia gas, axial electric field measurements have first been compared with a non-breakdown condition (without plasma) and in an 8 kV$_{pp}$ (with plasma), *ceteris paribus*. In closing, its evolution has been studied over three sets of experimental parameters: applied voltage, ammonia gas pressure, and the gap in between the dielectrics of the iDBD cell.

## 2. Experimental setup

The experimental setup is made of two main parts: an iDBD plasma reactor and the E-FISH diagnostic installation.

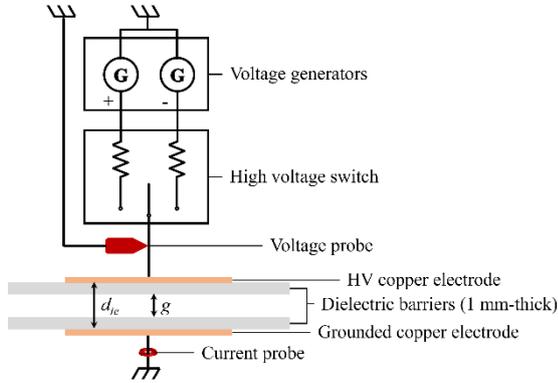

Figure 1: Electrical schematic of the iDBD.

Illustrated in Figure 1, a simplified scheme of the electrical setup used to perform the dielectric barrier discharge in impulse mode. The iDBD cell has a planar configuration, with two square shape copper electrodes (10 cm$^2$ surface area with a lateral dimension of 3.16 cm and a thickness of 80 μm). The upper electrode is connected to the high voltage supply, driving the discharge. The lower electrode is grounded. Two square shape dielectrics (5×5 cm²) of 1-mm thick, made of aluminum nitride, manufactured by Sceram Ceramics, are connected to each electrode, and separated by pure ammonia gas. These solid-state dielectrics exhibit a typical electrical resistivity higher than $10^{13}$ Ω×m and a typical dielectric constant at 1 MHz of 8. The gap distance $g$ between the two dielectrics can be modulated as desired. The distance in between the copper electrodes is denoted $d_{ie}$. The iDBD cell is positioned at the center of a 60 dm$^3$ grounded and closed 6-way cross spherical stainless-steel chamber (inner diameter 45 cm). This chamber allows to work under controlled atmosphere, thanks to a dedicated pumping system ranging from $10^{-4}$ Pa to $10^5$ Pa. The distance between the center of the iDBD cell and one of the quartz windows (Heraeus TSC-3® fused quartz) is 25 cm. In this study, the reactor was filled from secondary vacuum with pure ammonia gas (Air Liquide N50, 99.999% purity) at a pressure up to $2 \times 10^4$ Pa in static mode measured with a capacitive gauge (Pfeiffer CMR 361).

The iDBD plasma is generated with a homemade bipolar high-voltage power supply, which includes two direct high voltage generators (Technix SR15-R-1200) and a high-voltage high frequency switcher (Behlke HTS 301-15-SiC-GSM). This power unit is operated at 4 kHz, as in our previous work [9], with a symmetric square pulse high voltage and a rise time of about 1 μs. The high frequency switcher is triggered by an arbitrary function generator AFG3022C from Tektronix (bandwidth 25 MHz). Voltage and current measurements have been done by a CalTest Electronics CT4028 voltage probe (bandwidth from DC to 220 MHz) located on the high-voltage connector and a MagneLab CT1.0 current probe (bandwidth from 200 Hz to 500 MHz) located on the ground cable of the iDBD cell. Then, these electrical measurements are recorded by an LeCroy 104Xi oscilloscope (bandwidth from DC to 1 GHz and a 5 GS s$^{-1}$ rate) and averaged 20 times. To visualize the plasma, a picosecond high speed iCCD camera (Standford Computer Optics 4 Picos), equipped with an EF 100mm f/2.8 Canon macro lens, faces the iDBD cell. Set at 1 ns, the gate is synchronized with the discharge by using the same AFG that triggers the power supply. The resulting images is made by 5000 shots accumulated broadband emission from the discharge, with a temporal resolution of 1 ns.

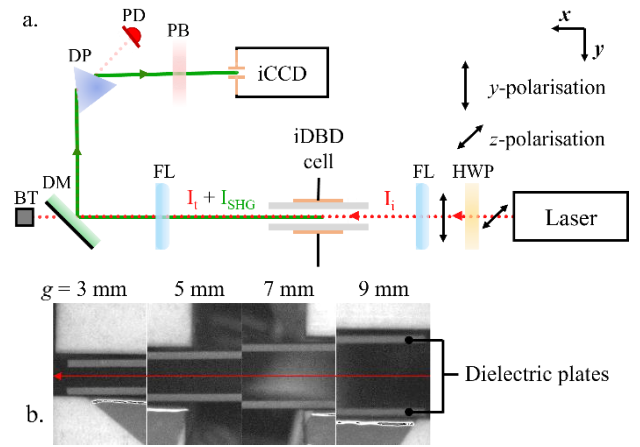

Figure 2: (a) Experimental scheme of the E-FISH optical bench for electric field strength measurement purpose. (b) The laser beam direction, represented by the red arrow, is always centered in the gap.

Shown in Figure 2, the E-FISH optical setup has been adapted from Chng and al. publication [35]. The fundamental output beam (1064 nm) of a Powerlite Precision II 8000 Nd:YAG laser (illustrated by the red dotted line) has a pulse duration of about 20 ns. To carry E-FISH measurements, the laser energy is attenuated to 60 mJ, measured before the quartz window of the iDBD reactor. The laser beam is polarized along the y-axis, parallel to the applied field, by a half-wave plate (HWP) illustrated by a double-headed arrow. The beam is focused on the center of the iDBD cell using the 500 mm plano-convex lens (FL). One can note that the distance from FL to the exterior surface of the quartz window is close to 26 cm. With a Gaussian beam profile, the Rayleigh length conditions an *x*-axis measurement resolution of approximately 3.5 mm with a beam waist estimated of about 70 µm in diameter. The second harmonic beam (illustrated by the green solid line) is generated collinearly with the fundamental pump beam. Then, the pump and the second harmonic beams are both spatially separated using a long pass dichroic mirror (DM) and a dispersive prism (DP). To monitor the stability of the laser, a Thorlabs DET10A2 photodiode (1 ns rise time) measures the intensity of the residual pump beam (illustrated by the light red dotted line) on the same oscilloscope used for the electrical measurements. In order to measure only the vertical component of the generated second harmonic, the beam is filtered by a polarizer (PB). Then it is focused on the entrance of an optical fiber which is connected to an optical spectrometer Jobin-Yvon Triax 550 equipped with a 1800 gr.mm$^{-1}$ grating and an i-Spectrum Two iCCD camera from Horiba Jobin Yvon. The integration time – the integration gate of the iCCD camera – is set at the minimum possible value i.e. 2 ns, temporally centered on the maximum of the laser pulse intensity, giving the temporal resolution with the best signal to noise ratio for our setup. To enhance more the signal to noise ratio, the second harmonic signal is accumulated from 100 discharges on the iCCD camera. One can note that in our conditions the discharges are stable in time, shape, and intensity. Finally, the accumulated acquisitions are averaged 20 times. Consequently, it is worth noting that each measured point of the E-FISH curves (including calibration ones) results from 2000 consecutive acquisitions. Given that these measurements take a relatively long time to carry out, special attention to the stability of the measurement conditions (room parameters, laser settings and measuring devices) must be systematically and necessarily taken into account. Furthermore, and to reduce the measurement uncertainty, especially for low measured intensity, the accumulated acquisitions is integrated from 531.4 to 533.6 nm.

To perform a time-resolved E-FISH diagnostic synchronized with the discharge, the laser pulse delay is controlled by the same function generator which operates the high voltage switch. Each E-FISH measurement is achieved with a 2 ns time steps. To enable cross-analysis between all the measurements, the speed of signal propagation in the power circuit (high voltage generators, switch, iDBD cell) and the measurement circuits are always taken into account in addition to the instrumental response time.

Great attention was paid to the synchronization between electrical and optical measurements, as E-FISH. Thus, both the synchronization (i.e. synchronous) and the phasing (i.e. temporal correction) of these measurements have been accurately executed, strictly considering the causal effects from Maxwell's laws, especially Maxwell-Ampère. The latter implies two elementary notions in this context: i) A displacement current cannot occur without local electric field variations, here within the iDBD cell; ii) An electric field can exist without conduction current, but a discharge current (mostly conduction) cannot exist without local drop of the electric field, here within the iDBD cell.

Table 1. Overview of the main parameters among the studied conditions

|  | Influence of | | |
| --- | --- | --- | --- |
|  | Voltage (sect. 3.3) | Pressure (sect. 3.4) | Gap (sect. 3.5) |
| $U$ (kV$_{pp}$) | 4 to 8 | 6 | 6 |
| $p$ ($\times 10^4$ Pa) | 1 | 0.8 to 2.0 | 1 |
| $g$ (mm) | 3 | 3 | 3 to 9 |
| $d_{ie}$ (mm) | 5 | 5 | 5 to 11 |

E-FISH diagnostic, electrical measurement and iCCD ultra-fast camera have been carried out for both, negative transition NT (from a higher to a lower voltage) and positive PT (from a lower to a higher voltage) of the upper high voltage electrode, for each condition listed in table 1. It can be noted that only the measurements obtained from NT are presented and discussed in this article. Indeed, although the measurements carried out are not perfectly identical, the overall behaviors observed on PT are similar to those of NT since the applied voltage pulses for PT an NT are symmetrical.

3. **Results and discussion**

To start with, the discussion will be focused on the calibration method, which is done prior to any E-FISH measurements. Then, the overall electrical characteristics are presented, followed by an electric field study with and without a discharge in pure ammonia. Finally, the influence of the applied voltage, the gas pressure and the gap distance $g$ will be presented and discussed.

## 3.1. E-FISH calibration

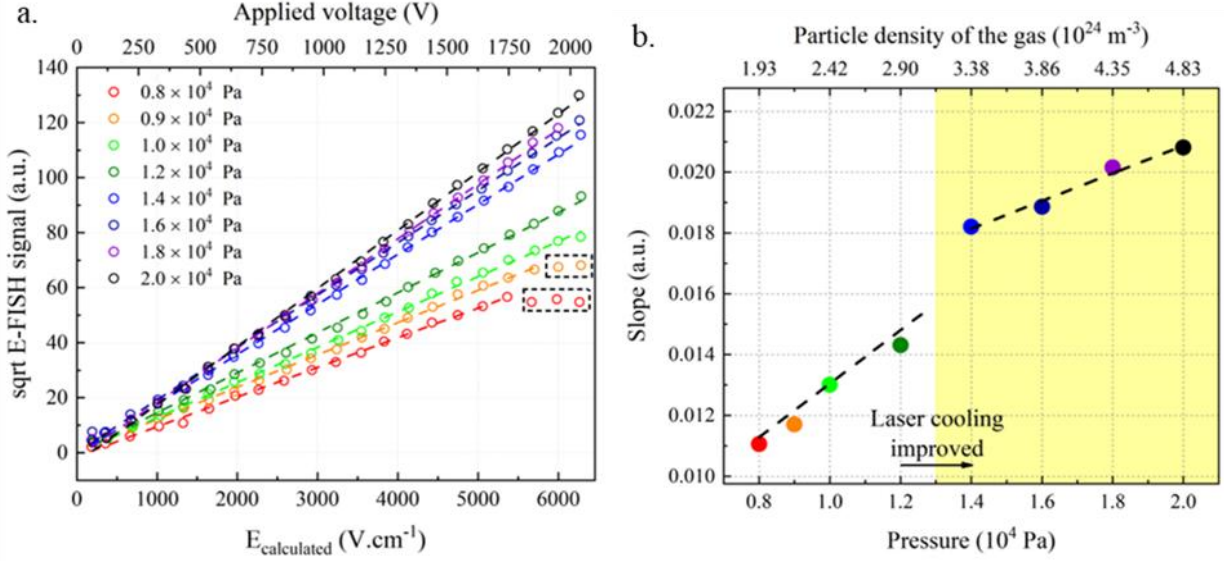

Figure 3: (a) The square root of the E-FISH signal as a function of the electric field calculated $E_{calculated}$ from DC applied voltages obtained for pressure from 0.8 to $2.0 \times 10^4$ Pa with a 3-mm gap $g$. Circles refer to the experimental data and dashed lines to a linear fit Dashed squares in figure 3a evidence diverging behavior from the calibration data curve trend. (b) The square root of the E-FISH signal slope, calculated from Figure 3a, as a function of the pressure and the particle density. The color scale remains the same.

The aim of calibrating the E-FISH measurements is twofold. First, it aims to investigate if there is a relationship between the second harmonic intensity and a supposedly known electric field. Second, the goal is to use the factors determined from these calibration curves in subsequent E-FISH measurements. It is important to note that all the calibration experiment is performed in the same experimental setup (*cf.* Figure 2) used for E-FISH measurements. This procedure is crucial since the second harmonic intensity could exhibit a dependency on both the electrode dimensions and the beam characteristics at the focal point. Additional information regarding the latter can be found in the work of Chng *et al.* [36].

From data calibrations, the square root of the E-FISH signal is plotted in Figure 3a as a function of $E_{calculated}$. The latter was calculated with DC applied voltages from square one (without previous breakdown), for all the studied conditions (applied voltage, pressure and gap). Therefore, the induced electric field is assumed to be uniform across the gas gap. Since, the AlN dielectrics are identical and due to the intrinsic symmetry of our iDBD cell, the value of $E_{calculated}$ along *y*-axis comes from Eq. 1 [37–41]:

$$E_{calculated} = U / \left( g + 2. d_D \cdot \frac{\varepsilon_g}{\varepsilon_D} \right) \quad \text{(Eq. 1)}$$

where $\varepsilon_g$ and $\varepsilon_D$ are respectively the permittivity of the ammonia gas (1.0062) and of the AlN dielectrics (8, given by the manufacturer), $d_D$ is the dielectric thickness, $g$ is the gap distance between the two dielectrics and $U$ is the applied voltage. All curves plotted in Figure 3a are done by increasing a positive applied voltage $U$. Overall, for all the conditions listed in Table 1, the square root of the second harmonic signal scales linearly with $E_{calculated}$ with a sensitivity threshold of approximately 150 V.cm$^{-1}$. All the linear curves (dashed lines) fit the description of a third order nonlinear process [29], using the following expression [32,33,35]:

$$I_{SHG} = A \cdot (N \cdot I_{pump} \cdot E_y)^2 \quad \text{(Eq. 2)}$$

where $I_{SHG}$ is the intensity of the induced second harmonic, $A$ is a measured calibration constant, $N$ is the gas density, $I_{pump}$ is the intensity of the pump laser beam and $E_y$ is the global electric field to be measured along the *y* axis, in accordance with our setup (*cf.* figure 2a). According to this Eq. 2, $I_{SHG}$ is proportional to the square of $E_y$. This implies that the only information that can be deduced is the modulus of the electric field. The direction of the measured electric field cannot be distinguished from one direction to its opposite. Some may conjecture it from other discharge parameters (*e.g.* current [41]). Note that all the following electric field values or curves discussed refer to the electric field strength along the y-axis, $E_y$.

One can notice some experimental data highlighted by dashed squares in Figure 3a. For these datasets, experimental points diverge from the linear regression, starting from 1850 V for the $0.8 \times 10^4$ Pa pressure condition and higher applied voltage (1950 V) for the $0.9 \times 10^4$ Pa one. These squared measurements could arise from a shielding effect, which might be due to the presence of charges accumulated on the dielectric surfaces in contact with the gas. The origin mechanism of this shielding effect without plasma is not clearly elucidated yet. Nevertheless, it seems that reaching the breakdown voltage condition at fixed pressure and gap, the direct applied voltage could generate accumulated charges on each dielectric surface, leading to the formation of a counter field. This counter field magnitude could be high enough to obtain a steady global $E_y$ while the applied voltage is increasing. Worth noticing that this phenomenon can arise for lower voltages when the pressure decreases, as it is well known regarding the Paschen's law and according to the ammonia Paschen curve given by Radford et al. [42].

To highlight the pressure effect on the second harmonic signal, the electric field calibration data plotted in Figure 3a, have been taken firstly for 0.8, 0.9, 1.0, 1.2, and secondly for 1.4, 1.6, 1.8 and 2.0 × $10^4$ Pa. Indeed, the two sets of data have been carried out with two different $I_{pump}$ values due to an improvement of our laser cooling system after the first dataset. The slope of the corresponding calibration curve has been plotted against the pressure and the theoretical particle density of the gas (calculated with ideal gas law), in Figure 3b. From this figure, a linear trend between the pressure and the square root of the second harmonic signal slope is observed for each dataset. This linear behavior is quite consistent with Eq. 2 where the square root of the second harmonic signal in a pure gas is proportional to $N$, the gas density, and obviously proportional to $I_{pump}$, the pump laser beam intensity.

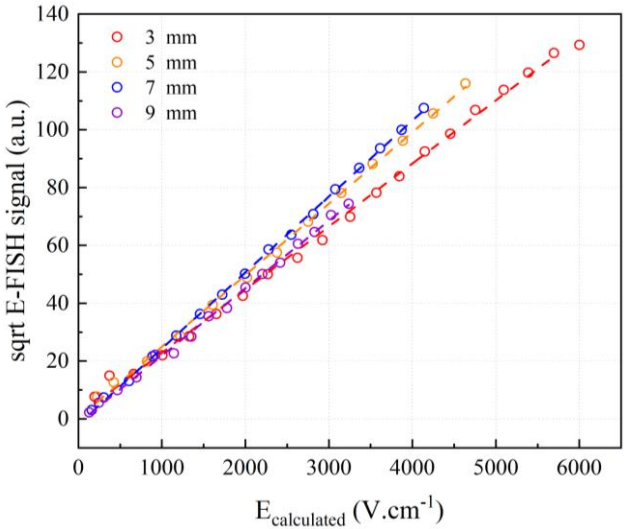

Figure 4: The square root of the E-FISH signal as a function of the electric field calculated $E_{calculated}$ from direct applied voltages obtained for pressure $10^4$ Pa, gap $g$ from 3 to 9 mm. Circles refer to the experimental data and dashed lines to a linear fit.

Calibration data have also been measured for 3-, 5-, 7- and 9-mm gap $g$, gathered in figure 4. By varying the gap $g$ from 3 to 9 mm, the calibration data slopes are alike.

This tendency is in good accordance with Eq. 2, where the square root of the second harmonic signal is not influenced by the gap. Moreover, our previously stated assumption of a uniform electric field across the gas gap seems valid in this gap range.

However, one can note that the slope values of the linear regressions associated to each curve are slightly different to one another (± 10 % max). As discussed earlier in figure 3b, these results obtained before the cooling improvement show again how impactful could be the stability of the laser intensity when carrying out E-FISH measurements. In our case study, it therefore appears necessary to carry out calibration curves for each experimental condition studied. By doing so, this methodology ensures a high confidence level in the $E_y$ strength measurements obtained with the E-FISH diagnostic: they are preceded and followed by their own calibration, in all cases. In addition by following this procedure, the contribution of relative humidity and room temperature variations from one day to another is also always taken into account.

### 3.2. Electrical and E-FISH characteristics of the iDBD cell

The aim of this section is to analyze the E-FISH measurements qualitatively without and with a plasma. To extend the discussion in the sections dealing with parametric studies (3.3, 3.4 and 3.5), electrical and E-FISH characteristics of the iDBD cell are detailed here, giving an in-depth knowledge of the latter.

Figure 5 exhibits the plot of typical electrical variations over time, measured in pure ammonia pulsed discharge at $10^4$ Pa, with 6 kV$_{pp}$ applied and a 3 mm gap. Three temporal aspects can be observed. Figure 5a brings out the periodicity of the voltage changes. A pulsed plasma ignition occurs every half-period with $T_{/2}$ = 125 μs, while the voltage at the upper electrode varies from its highest (+ 3 kV) to its lowest value (- 3 kV), defined as the negative transition NT, and from its lowest (- 3 kV) to its highest (+ 3 kV) value, defined as the positive one PT. The resulting periodic current of these voltage transitions tends to classify this iDBD regime as a homogeneous discharge contrary to a filamentary discharge characterized by a series of intense current strikes [8,12,43,44], as in sine-AC regime. A second temporal aspect concerns the time needed for the voltage to reach its set value, at a transition. For both NT and PT, the applied voltage reaches more than 95 % of its set value in approximately 1 μs, corresponding to capacitor charge regime of the global system (power supply and iDBD cell) reaching 3τ. The third and last temporal aspect concerns the discharge duration. From Figure 5b, the major current variation due to the discharge is measured for about 100 ns with a rise time less than 10 ns.

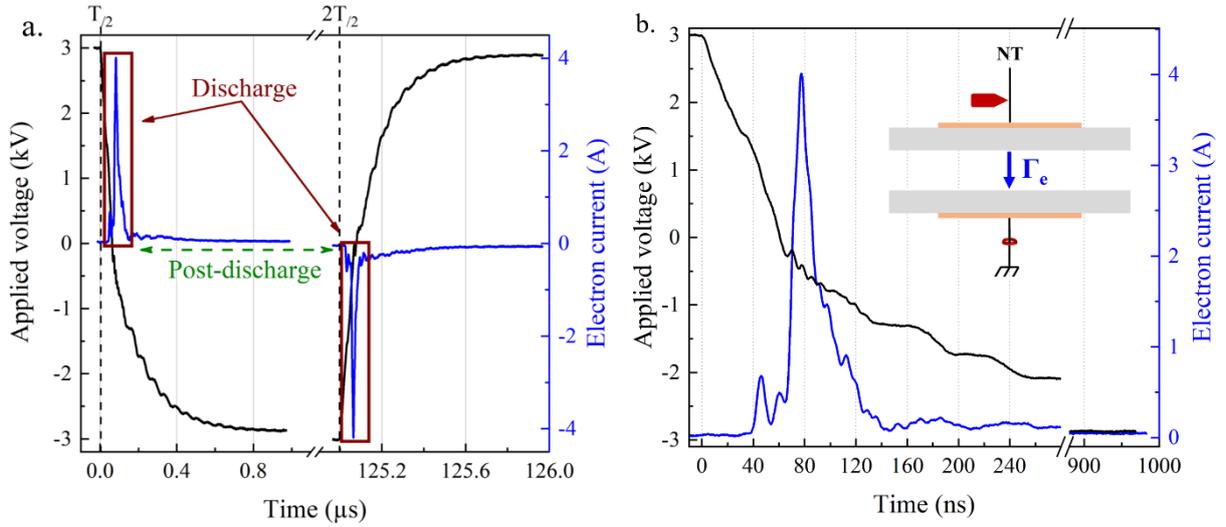

Figure 5: Time-resolved electrical characteristics (voltage and current), measured in a 6 kV$_{pp}$ squared discharge in ammonia at $p = 10^4$ Pa, $g = 3$ mm, (a) with a break from 1 to 125 μs for two consecutive transitions (NT and PT) and (b) a focus on the negative transition (NT) during the first 1 μs, with a break from 280 to 880 ns.

According to Langmuir waves formula, the plasma oscillations period $T_j$ of a charged particles $j$ is defined as:

$$T_j = \frac{2\pi}{\omega_{pj}} \quad \text{with} \quad \omega_{pj} = \sqrt{\frac{n_j q^2}{m_j \varepsilon_0}} \quad \text{(Eq. 3)}$$

Where $m_j$ is the mass, $n_j$ the number density and $\omega_{pj}$ plasma oscillation of the considered species ($j = i$ or $e$, respectively for ions or electrons), $q$ the electric charge and $\varepsilon_0$ the permittivity of free space. On these time scales, ions can reasonably be considered slow or immobile compared to electrons since $m_i$ is much higher than $m_e$, typically more than $1.836 \times 10^3$ heavier considering the lightest ion, a proton. Therefore, the measured current during this third temporal aspect is mainly due to the dynamics of the electrons of the plasma discharge and is then referred to as an electron current in our conditions. One can note that a positive measured current presented in Figure 5b means that the electron flux $\Gamma_e$ is oriented from the upper dielectric surface to the lower one, as indicated in the insert.

E-FISH measurements have been carried out and synchronized with electrical measurements. Figure 6 shows the electrical characteristics and the associated electric field temporal variations measured at the center of the dielectric gap for a non-plasma condition (applied impulse voltage 3 kV$_{pp}$) and for a pure ammonia plasma discharge (applied voltage 8 kV$_{pp}$), during NT and up to 550 ns.

From Figure. 6a, the maximum electric field value is about 4600 V.cm$^{-1}$ while a voltage of +1.5 kV is applied. One can note that the latter was constant during the last 124 μs, corresponding to the previous $T_{/2}$, here the end of PT. As expected, this electric field value found is very closed to the theoretical one at 4613 V.cm$^{-1}$, given by Eq. 1. At $t_0 = 0$ ns, the negative transition starts (voltage inversion at the upper cathode) and the electric field slightly increases before dropping to a minimal value of about 200 V.cm$^{-1}$ when the applied voltage reached 0 V, thus not so far from the sensitivity threshold of our setup. At this stage, it should be noted that the value of the electric field within the iDBD cell is certainly lower than that given by the $E_y$ measurement. Following this electric field drop, a slight variation in electron current is measured, reaching a maximum value of 0.5 A. Since the electrical parameters applied in this case do not satisfy the conditions required to initiate a discharge, these current variations could be associated to variations of the electric field throughout the power supply circuit of the iDBD cell, including both copper electrodes.

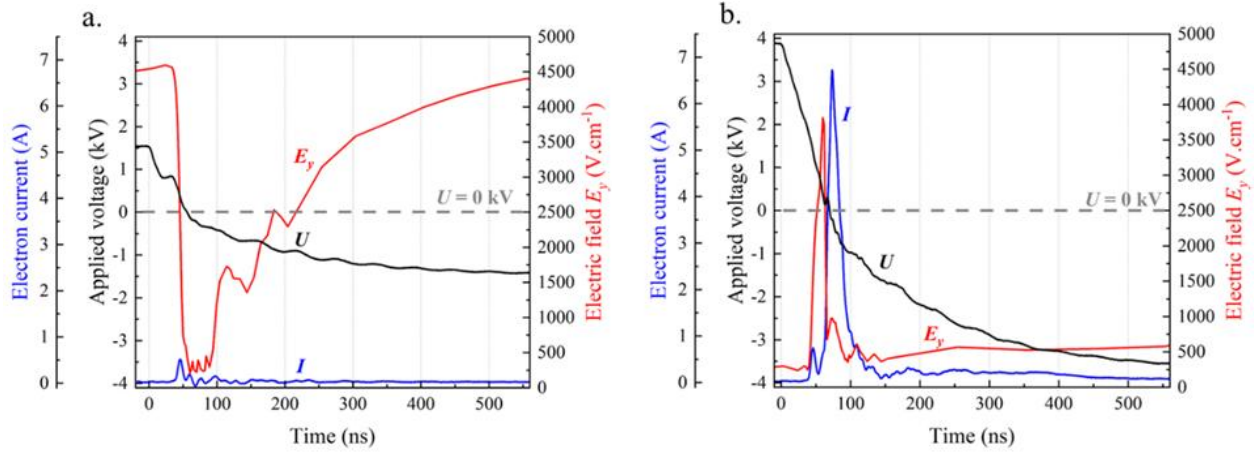

Figure 6: Electric field $E_y$ (red curve), voltage and current (black and blue curves) variations measured (a) without plasma at 3 kV$_{pp}$ and (b) with plasma at 8 kV$_{pp}$, in pure ammonia, $p = 10^4$ Pa and a 3-mm $g$.

This last assertion will be developed and discussed in the section 3.3. Beyond 70 ns, the applied voltage gradually increases to reach the − 1.5 kV set value meanwhile the electric field tends to approach the initial value of about 4600 V.cm$^{-1}$ over around 500 ns. The variations of the electric field strength and the voltage describe altogether the behavior of a classical capacitor.

From Figure 6b, the current drastically increases up to 6.5 A during the NT since the ignition conditions to generate an impulse plasma is fulfilled. One can note that after the beginning of the NT the electric field strength is minimal at about 250 V.cm$^{-1}$ before 40 ns and about 600 V.cm$^{-1}$ beyond 500 ns, whereas the applied voltage amplitude is maximal at 4 kV. This behavior is at the opposite of the one previously described, *i.e.* the classical capacitor. Furthermore, the main variations of the electric field strength start at the same time as the first current fluctuation (at 40 ns) with a maximum $E_y$ value of about 3845 V.cm$^{-1}$, reached at 60 ns. One can note that this maximum value of $E_y$ is lower than the previous maximum of 4600 V.cm$^{-1}$ measured in case of classical capacitor behavior, whereas the applied voltage is much higher than 1.5 kV. Indeed, the discharge ignition occurs only 60 ns after the beginning of NT. As soon as the current increases rapidly, resulting to a drastic drop of $E_y$ at the same time. It is interesting to note that this temporal profile of $E_y$ can be found in previous study while the experimental setup is different [22,35,45,46].

Overall, these results obtained with breakdown condition seem to highlight once more the existence of a counter electric field $E_{cf}$ explaining, on one hand the maximal value of about 3845 V.cm$^{-1}$ prior to the discharge ignition in these conditions and on the other hand the low $E_y$ values measured before and after the discharge during the NT (similarly during the PT). This latter seems to unveil a counter-field $E_{cf}$ whose variations oppose those induced by the voltage applied to the electrodes, lasting the entire iDBD regime.

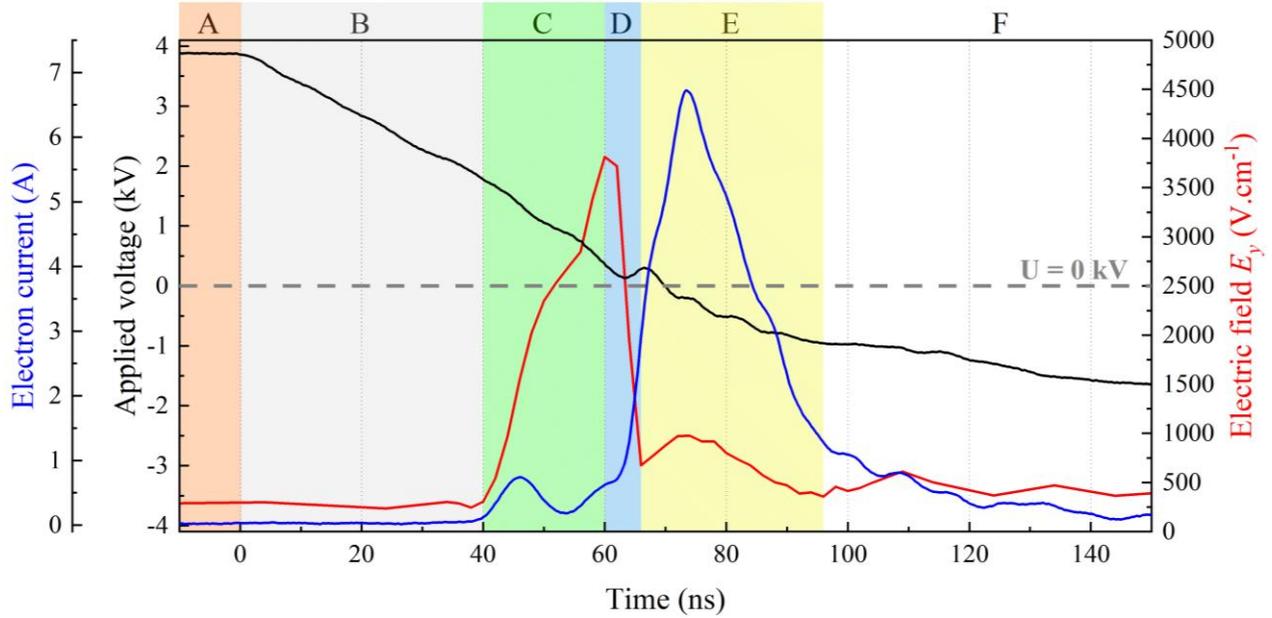

Figure 7: Scale time expansion around the electric field $E_y$, applied voltage and current main variations of figure 6b. Pure ammonia, pressure $10^4$ Pa, 3-mm $g$.

From figure 6b, the figure 7 displays a scale time expansion of the beginning of NT. The variations of $E_y$ during the iDBD regime, as for voltage and current, have been discriminated into 6 domains, labelled from A to F. Domain A is the steady state domain, where the $E_y$, the applied voltage and the current remain constant. Domain B is bounded by the start of the transition ($t_0$) and the first steep variations of current and $E_y$, which taken together correspond to the dielectric relaxation [47–49] response of ideal dielectric dipoles to an alternating external electric field, here of about 40 ns for the aluminum nitride in our conditions. Domain C relates to the fast increase of the electric field to its maximum value. Domain D relates to the discharge ignition, which leads to the fall of $E_y$. Domain E is the electric field strength measurement during the discharge. Finally, domain F pertains to post-discharge events still the next half period, in this case with domain A for the PT.

### 3.3. Influence of the applied voltage

As mentioned before, the physical mechanism behaviors between the PT and NT remain the same and only results associated to the NT are described here.

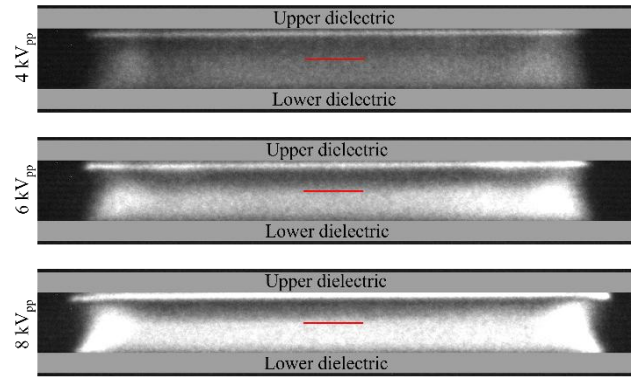

Figure 8: iCCD ultra-fast camera images of iDBD discharge for three applied voltage conditions: 4, 6 and 8 kV$_{pp}$. These images were taken at maximum discharge current during the NT with a 1 ns integration gate. Red line evidences the estimated E-FISH probed volume. Pure ammonia, pressure $10^4$ Pa, 3-mm $g$.

Figure 8 displays three iCCD ultra-fast camera images obtained from 5000 shots accumulated emission. They show that the discharge is quasi-homogeneous along the $x$-axis whatever the applied voltage condition at $10^4$ Pa and a 3-mm g. One can note that the main E-FISH signal come from a tiny probed volume estimated here with laser beam waist around 70 μm in diameter and 2-3 mm length [29,35] along the $x$-axis. The discharge can be considered homogeneous in this probe volume.

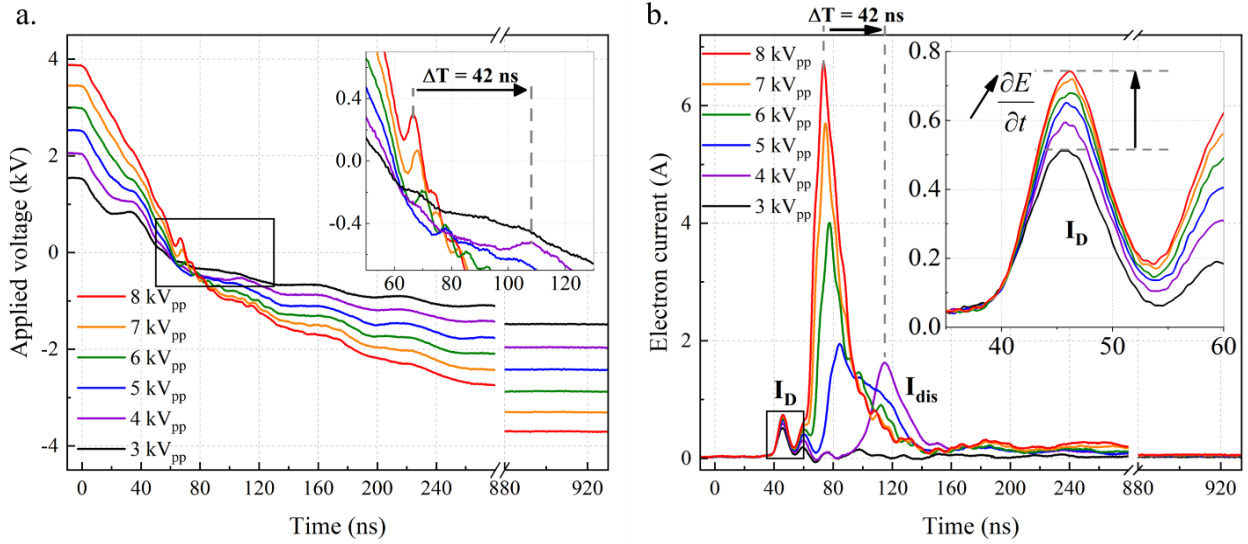

Figure 9: Temporal evolution of (a) the applied voltage and (b) the related electron current for the five studied conditions, from 4 kV$_{pp}$ (purple curve) to 8 kV$_{pp}$ (red curve) with plasma and the 3 kV$_{pp}$ without plasma (black curve), during NT. Pure ammonia, pressure $10^4$ Pa, 3-mm $g$.

To study the applied voltage influence on $E_y$, five electrical conditions have been investigated from 4 to 8 kV$_{pp}$, *ceteris paribus*. Figure 9a introduces the temporal evolution of the applied voltage during the NT for each condition. One can note that the 3 kV$_{pp}$ condition is also added in figure 9, as a non-plasma reference.

From Figure 9a, two main characteristics can be observed. The first one is the initial voltage slope variation of each curve. As the applied voltage is higher, the voltage slope variation steepens from about − 35 to − 56 V.ns$^{-1}$ between $t_0$ and $t_0$+60 ns. The second one addresses the short variations (~10 ns) which occur between $t_0$+60 and $t_0$+120 ns depending on the initial applied voltage, as shown in the insert. They are associated to the domain E of each current curve of Figure 9b. One can note that the power of our high voltage generators is intrinsically limited (here 1200 W). Hence when the discharge occurs, the generator involved in the NT is not able to keep a standard evolution of the voltage, as observed in the non-plasma case (*e.g.* at 3 kV$_{pp}$). This occurrence acts as good indicator of the pulse plasma ignition in our setup. Indeed, Figure 9b displays a same time shift $\Delta T$ of 42 ns between the 8 and 4 kV$_{pp}$ conditions for the maximum measured currents. Moreover, as expected, the maximum currents decrease from 8 to 4 kV$_{pp}$ conditions. However, this decrease in maximum current is delayed by $\Delta T$. From $t_0$, $\Delta T$ is dependent to ignition conditions, which are reached at longer time scale for lower voltage. This ignition conditions will be discussed in the following part.

The insert in Figure 9b focuses on the first part of the measured current which is time-invariant (4 to 8 kV$_{pp}$). Even in the case of a non-plasma condition (3 kV$_{pp}$), this time-invariant current also remains. Thus, it seems independent and distinct from the discharge. The first current amplitude measured on it increases gradually for higher applied voltage, from 0.5 to more than 0.7 A for 3 to 8 kV$_{pp}$, respectively. This time-invariant current which occurs during our measurements can be described by Maxwell-Ampere's law (Eq.3), stated as follows:

$$\overrightarrow{rot}\boldsymbol{B} = \mu_0\varepsilon_0\frac{\partial \boldsymbol{E}}{\partial t} + \mu_0\boldsymbol{j}_C = \mu_0(\boldsymbol{j}_D + \boldsymbol{j}_C) \quad \text{(Eq. 3)}$$

Where $\boldsymbol{B}$ is the magnetic field vector, $\boldsymbol{E}$ is the electric field vector, $\boldsymbol{j}_D$ is the displacement current density vector due to temporal electric field variations, $\boldsymbol{j}_C$ is the conduction current density vector due to charge flow, $\varepsilon_0$ and $\mu_0$ are respectively the permittivity and the permeability of free space.

Thenceforth, the time-invariant current $I_D$ (*cf.* Figure 9b), corresponds to the first term of Eq. 3, which describes the displacement current density vector $\boldsymbol{j}_D$. Proportional to $|\varepsilon_0 \partial\boldsymbol{E}/\partial t|$ and thus link to the voltage slope variations, the amplitude of the displacement current $I_D$ is relative to time-varying electric field within a dielectric/capacitive system, *i.e.* the iDBD cell. By adding previous observations and discussions of Figures 6 and 7, the current $I_D$ is induced by any change in electric field of a system involving a capacitive effect. In our case, while the NT starts at $t_0$, this first variation of $I_D$ occurs only at 40 ns because of the dielectric relaxation response as described by the domain B. Afterwards, as soon as the breakdown occurs, the ammonia gas medium behaves more like a conductor. Consequently, the main contribution to the measured current $I_{dis}$ of the domain E is the conduction current $I_C$, corresponding to the second term of Eq. 3.

The results obtained via E-FISH measurements on $E_y$ and associated current measurements have been compiled for the five voltage conditions mentioned above, plus $E_y$ measurements for 5.5 kV$_{pp}$. They are presented in Figure 10a and 10b, respectively.

In Figure 10a, one can observe at the beginning of NT (domain B, as for A) that whatever is the applied voltage, a very low and quasi-constant value of the $E_y$ strength is measured (~ 250 V.cm$^{-1}$). This means that a counter-field $E_{cf}$ seems to adjust to the applied voltage. Furthermore, as shown in figure 10a, the first fast variations of $E_y$ (domain C) seem to begin always at the timescale whatever is the applied voltage, at about 40 ns after $t_0$. As discussed above, this time-invariant behavior is also observed in the figure 10b with the first measurement of the displacement current $I_D$, indicated by the black arrow. One can note that both $E_y$ maximum and the beginning of $I_{dis}$ measurements arise at a time scale ranging from 60 to 70 ns, corresponding to the end of domain C. Moreover, the transition to domain D, *i.e.* the $E_y$ fast drop, appears earlier, from about $t_0$+70 to $t_0$+65 ns with the increase of applied voltage from 5 to 8 kV$_{pp}$. Altogether, the electric field slope in domain C is quasi-invariant and thus comforts again the existence of the counter-field $E_{cf}$, which adjusts to the applied voltage. Indeed, $E_{cf}$ can only be linked to the charge state of the iDBD cell induced by the previous discharge, here during PT. Concerning the 4 kV$_{pp}$ condition, the $E_y$ drastic fall appears only at about $t_0$+100 ns. The latter is also in good agreement with the time shift $\Delta T$ previously mentioned between the 8 and 4 kV$_{pp}$ conditions (*cf.* figure 9).

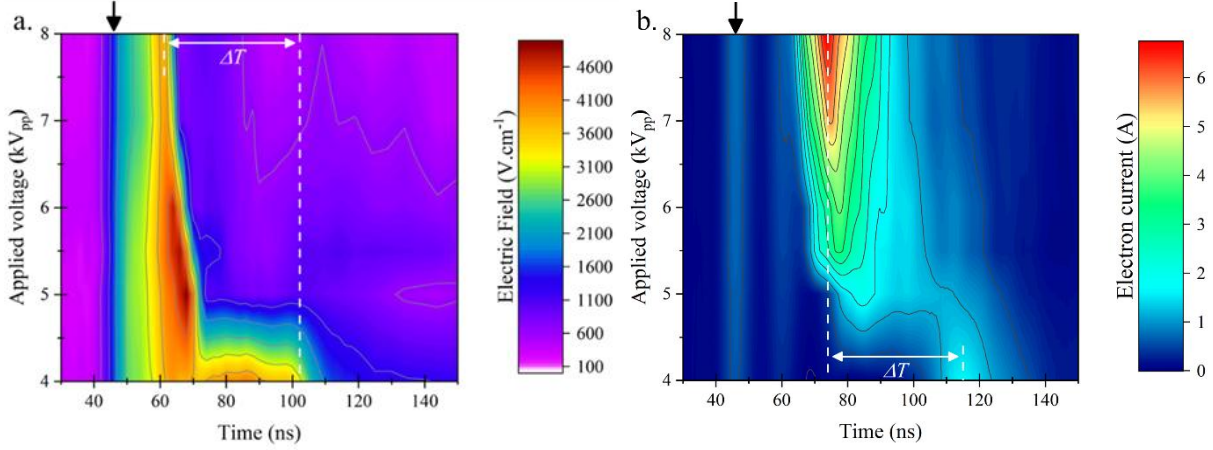

Figure 10: 2-D representations over time and applied voltage of (a) the E-FISH measurements of $E_y$ and (b) the measured current for the five studied voltage conditions, *i.e.* from 4 to 8 kV$_{pp}$, during NT. Pure ammonia, pressure 10$^4$ Pa, 3-mm *g*.

In Figures 10a and 10b, one can note that the 4 kV$_{pp}$ condition exhibits a particular behavior compared to higher voltage, with an $E_y$ of about 3800 V.cm$^{-1}$ during 40 ns, up to $t_0$+100 ns approximately. This behavior is due to the threshold ignition condition, which consequently requires a long time for its establishment. For higher applied voltages, an $E_y$ maximum measured value in domain C is reached between 5 and 6 kV$_{pp}$ with a value of about 4900 V.cm$^{-1}$. Above 6 kV$_{pp}$, it decreases whereas the measured current increases up to 6.7 A for 8 kV$_{pp}$, as shown in Figure 10b. Altogether, these results indicate that at the end of each transition NT and PT there are charges in the gas, which favors more or less the next discharge ignition. Indeed, at the transition from domain C to domain D, both the decrease of $E_y$ maximum measured value from 5 to 8 kV$_{pp}$ and the earlier drastic fall of about 5 ns, allow to evidence a better conductivity of the gas volume, *i.e.* more remaining charges in gas volume during domains A and B.

### 3.4. Influence of the pressure

Following the same approach as used previously, an investigation of the influence of gas pressure has been conducted within the range of 0.8 to 2.0 × 10$^4$ Pa, while keeping other parameters constant (*i.e.* 6 kV$_{pp}$ and a 3-mm gas gap).

To begin with, Figure 11 presents four images captured using the same procedure as described previously, during NT discharge, corresponding to 0.8, 1.0, 1.6 and 2.0 × 10$^4$ Pa conditions from top to bottom. These images are not directly comparable due to the inability to maintain a constant iCCD gain across the range of gas pressures studied. The iDBD plasma appears quasi-homogeneous in the probed volume even if some columnar structures appear in case of 2.0 × 10$^4$ Pa. The columnar-like aspect suggests a spatial reorganization of the discharge itself to favor the ignition and sustainment of the plasma [50].

Figure 12 shows time-resolved 2-D representations over time and gas pressure of the E-FISH measurements of $E_y$ and the measured current for the seven studied pressure conditions, from 0.8 to 2.0 × 10$^4$ Pa, during NT. The time-varying electric field exhibits some dominant trends.

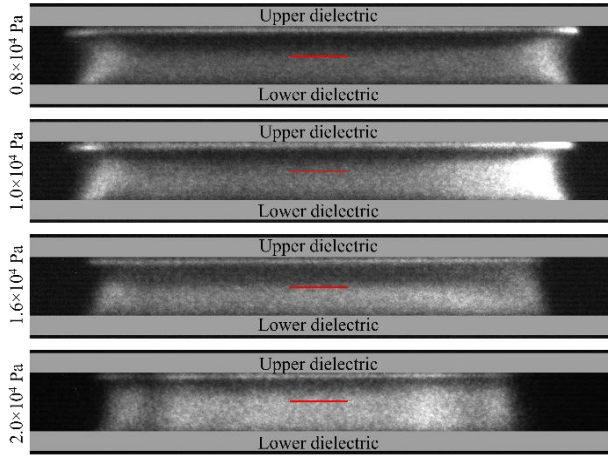

Figure 11: iCCD ultra-fast camera images of iDBD discharge for four gas pressure conditions: 0.8, 1.0, 1.6 and $2.0 \times 10^4$ Pa. These images were taken at maximum discharge current during the NT with a 1 ns integration gate. Red line evidences the estimated E-FISH probed volume. Pure ammonia, applied voltage 6 $kV_{pp}$, gap $g$ 3 mm.

Regardless of the gas pressure, the two first vertical iso-value lines at $t_0+42$ and $t_0+45$ ns denote that the electric field increases in domain C in the same way with a slope of about 210 V.cm$^{-1}$.ns$^{-1}$. Besides, the $E_y$ maximum value increases as the gas pressure is increased in the iDBD cell, from 4.15 to 5.10 kV.cm$^{-1}$. By dividing the obtained $E_y$ data in function of the gas density, the reduced electric field could be obtained as shown in Figure 13. The highest reduced electric field ( ~ 215 Td) is obtained for the lowest pressure (p = $0.8 \times 10^4$ Pa), as described in Chng and al. work [35]. One can note that the higher the pressure, the later the field strength maximum value is reached. Furthermore, the $E_y$ maximum measured value decreases as the pressure increases. Therefore, the time needed to establish discharge increases with pressure $p$. These results show that the discharge establishment follows a specific behavior. This corresponds to the increasing part of the classical Paschen's curve as a function of $p \times g$.

Furthermore, from the vertical iso-value lines at $t_0+45$ and $t_0+60$ ns in Figure 12b, the two well defined displacement currents $I_D$ are invariant in time and in amplitude whatever is $p$. This behavior was expected since at a fixed applied voltage, the temporal voltage variation is strictly similar before the ignition of the discharge. The latter confirms that without plasma the pressure has no effect on the iDBD cell's electrical characteristics (domain C).

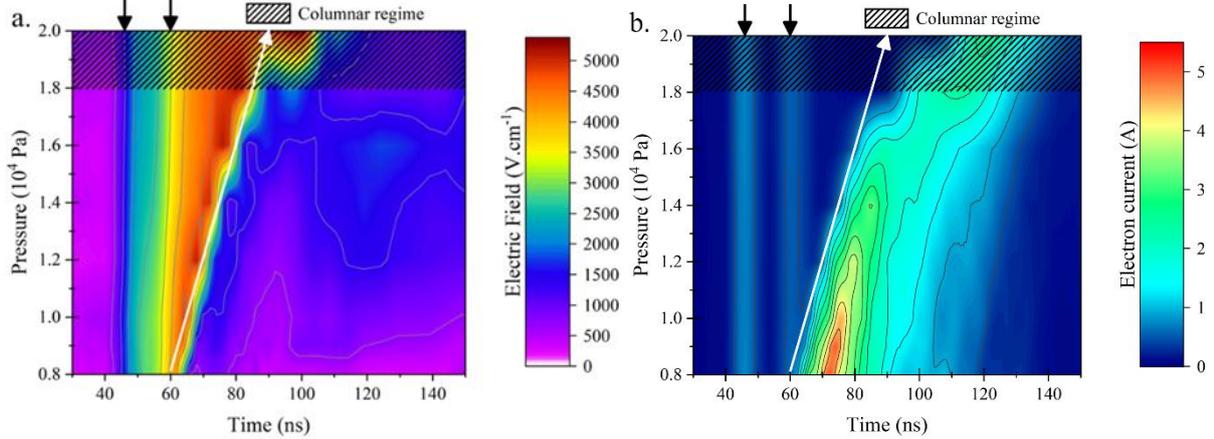

Figure 12: 2-D representations over time and gas pressure of (a) the E-FISH measurements of $E_y$ and (b) the measured current for the seven studied pressure conditions, from 0.8 to $2.0 \times 10^4$ Pa, during NT. Hatched area indicates conditions where the discharge has a columnar aspect. Pure ammonia, applied voltage 6 $kV_{pp}$, 3-mm $g$.

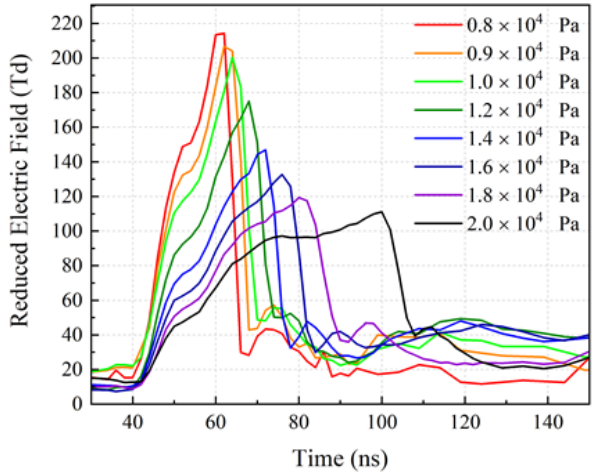

Figure 13: Evolution of the reduced electric field over time. Pure ammonia, applied voltage 6 kV$_{pp}$, 3-mm $g$.

Moreover, regarding the evolution of the discharge current $I_{dis}$ maximum (in domain E), it is the opposite behavior of the $E_y$ maximum reached at the beginning of domain D. Indeed, the $I_{dis}$ maximum value decreases by a factor of about 2 whereas the $E_y$ maximum value increases by a factor of about 1.2, between 0.8 and 2.0 × 10$^4$ Pa. Nevertheless, this behavior on $I_{dis}$ maximum is consistent with the one described by the reduced electric fields.

### 3.4. Influence of the gap

This section discusses the results of the gap study, ranging from 3 to 9 mm with other parameters remaining constant (*i.e.* 6 kV$_{pp}$ and 1.0 × 10$^4$ Pa of pure ammonia). As previously mentioned, only the results related to NT are presented and the output beam remains centered on the inter-dielectric gap.

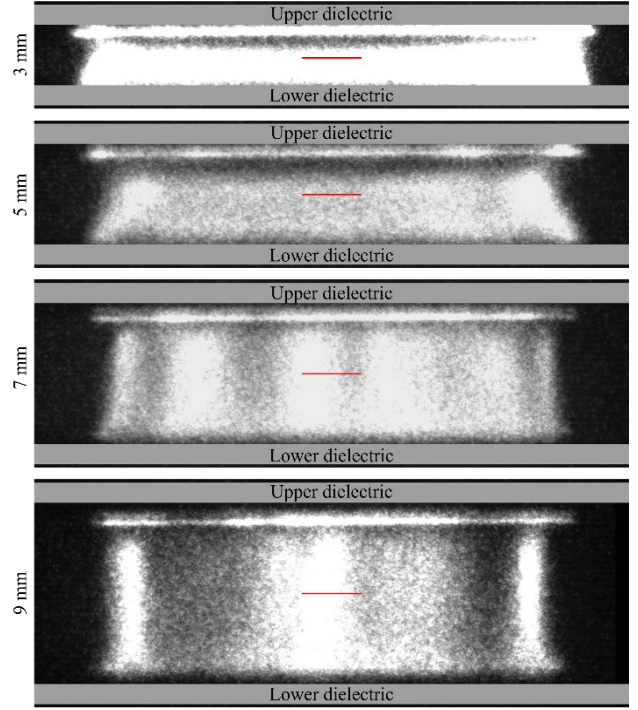

Figure 14: iCCD ultra-fast camera images of iDBD discharge for four gap conditions: 3, 5, 7 and 9 mm. These images were taken at maximum discharge current during the NT with a 1 ns integration gate. Red line evidences the estimated E-FISH probed volume. Pure ammonia, applied voltage 6 kV$_{pp}$, pressure 1.0 × 10$^4$ Pa.

Unlike the voltage and pressure parameters, the variation of the gap $g$ intrinsically modifies the capacitive properties of the iDBD cell. Consequently, slight variations of $g$ can lead to large modifications in the homogeneity of the discharge along the *x*-axis, as shown in the Figure 14. These modifications are well observed in our conditions with a 7-mm $g$. Note that in the vicinity of the volume probed, the columnar structures are stationary.

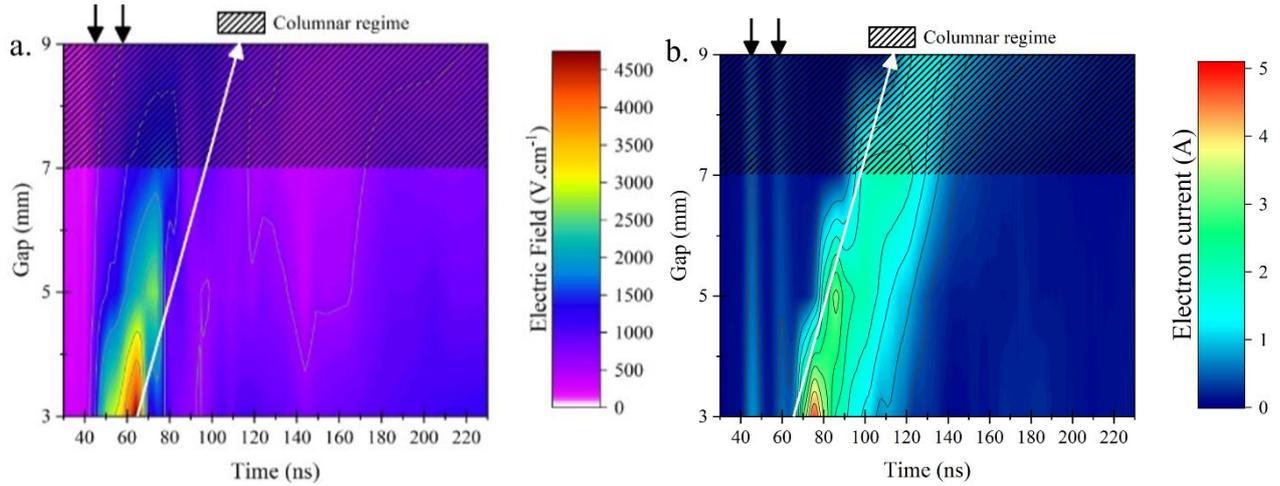

Figure 15: 2-D representations over time and gap $g$ of (a) the E-FISH measurements of $E_y$ and (b) the measured current for the four studied gap conditions, from 3 to 9 mm, during NT. Hatched area indicates conditions where the discharge has columnar aspects. Pure ammonia, applied voltage 6 kV$_{pp}$, pressure $1.0 \times 10^4$ Pa.

The E-FISH measurements of $E_y$ and associated current are displayed for these four gap conditions in Figure 15a and 15b, respectively. In Figure 15a, one can observe that at the beginning of NT (from domain A to B), whatever is the gap $g$, a very low and quasi-constant value of the $E_y$ strength is measured, closed to 250 V.cm$^{-1}$, similarly to the previous studied parameters, *i.e.* applied voltage and gas pressure. Once again, this result supports that the counter-field $E_{cf}$ adjusts itself to the discharge conditions. The Figure 15b gives the first variations of current $I_D$ at 40 ns, as previously, while the distance $g$ between the two dielectrics is significantly modified by a factor of 3. This confirms for any case that domain B, corresponding to the dielectric relaxation response of our setup, exclusively depending on the nature of the solid-state dielectrics (AlN) in our case. Obviously and regarding Eq. 3, the current $I_D$ decreases with the increase of $g$, for a given applied voltage (here 6 kVpp). As expected, by increasing the gap between the electrodes at constant applied voltage, the $E_y$ maximum measured value at the gap center decreases, from about 4600 to 950 V.cm$^{-1}$ with gap $g$ from 3 to 9 mm, respectively. With a low value of $E_y$ less than 1000 V.cm$^{-1}$ and a gap of 9 mm, the discharge can still be ignited compared to the higher $E_y$ values reached in domain C for the two previous parametric studies. Obviously, by increasing the gap, the $E_y$ measurements carried out at the center of $g$ moves mechanically away from the copper-dielectric electrodes. Therefore, this measurement location cannot correspond to the one where the electric field is maximum, causing the discharge ignition. Thenceforward, taking into account the small size of the volume probed, the measurement of $E_y$ is a local measurement, part of the electric field distribution in the gas. This last result supports once again the existence of persistent charges in the volume from one transition to the next one, PT to NT discussed here.

**Conclusion**

This work reports time-resolved electric field, by means of E-FISH diagnostic synchronized with electrical measurements (voltage and current) of an impulse dielectric barrier discharge iDBD in pure ammonia for the first time. Using these diagnostic tools, the iDBD have been qualitatively studied regarding different working conditions, as the applied voltage (3 to 8 kV$_{pp}$), the pressure $p$ (0.8 to $2.0 \times 10^4$ Pa), and the inter-dielectric distance, named gap $g$ (3 to 9 mm).

The sub-microsecond transitions from a steady-state voltage to another lead to the ignition of the plasma with a fast pulsed discharge current, which follows the Maxwell-Ampere's law. To match the ns-timescale phenomena, the E-FISH acquisition method has been improved, reaching a 2 ns temporal resolution with a 70 μm × 3.5 mm probed volume. The electric field $E_y$ variation without plasma emphases a classical capacitor behavior while the one obtained with plasma is completely different. Indeed, the low $E_y$ value measured before and after the discharge reveals the presence of a counter-field $E_{cf}$, which is opposed the one resulting from the applied voltage. More than that, this $E_{cf}$ seems to adapt itself to the discharge conditions. Synchronized measurements of applied voltage, current and $E_y$ enabled a complete description of the discharge and the post-discharge of iDBD, providing 6 defined temporal domains (A to F). Moreover, the electric field $E_y$ always reaches a maximum value before drastically dropping when the discharge current $I_{dis}$ starts to increase. However, the voltage influence study points out a threshold voltage condition beyond which the $E_y$ maximum value measured lowers while the $I_{dis}$ maximum value continues to increase. This competitive evolution of both quantities evidences the increase of persistent charges in volume for higher voltage. As expected for the pressure effect, both the reduced electric field and the $I_{dis}$ increase as the pressure decreases in accordance with Paschen's law. Ultimately, the $E_y$ measurements carried out at the center of the iDBD cell for different gap $g$ support the idea that a significant electric field distribution exists all along the 6 mentioned domains.
To characterize this electric field distribution, a specific E-FISH study could be carried out at various locations in the gap of the iDBD cell

**Acknowledgement**

This work has been funded by the French Agence Nationale de la Recherche (ANR) under project SYNERGY (ANR-20-CE05-0013).